\documentclass[aps,prc,twocolumn,showpacs,superscriptaddress]{revtex4}
\usepackage[english]{babel}
\usepackage{amsmath}
\usepackage{amsfonts}
\usepackage{amssymb}
\usepackage{graphicx}

\begin{document}

\title{Internal conversion of the low energy $^{229m}$Th isomer in the Thorium anion}

\author{E.~V.~Tkalya}
\email{tkalya_e@lebedev.ru}

\affiliation{P.N. Lebedev Physical Institute of the Russian
Academy of Sciences, 119991, 53 Leninskiy pr., Moscow, Russia}

\affiliation{National Research Nuclear University MEPhI, 115409,
Kashirskoe shosse 31, Moscow, Russia}

\affiliation{Nuclear Safety Institute of RAS, Bol'shaya Tulskaya
52, Moscow 115191, Russia}

\author{R.~Si}

\affiliation{Shanghai EBIT Lab, Key Laboratory of Nuclear Physics
and Ion-beam Application, Institute of Modern Physics, Department
of Nuclear Science and Technology, Fudan University, Shanghai
200433, Peoples Republic of China}

\affiliation{Spectroscopy, Quantum Chemistry and Atmospheric
Remote Sensing (SQUARES), CP160/09, Universit'e libre de
Bruxelles, Av. F.D. Roosevelt 50, 1050 Brussels, Belgium}

\date{\today}

\begin{abstract}
A process of the decay of the anomalously low lying nuclear isomer
$^{229m}$Th$(3/2^+,8.28 \pm 0.17$~eV) in the Thorium anion
(Th$^-$) via the internal conversion (IC) channel is studied. We
show that the half life of the nuclear isomer in the
$6d_{3/2}^37s_{1/2}^2$ ground state and in the $6d_{3/2}^2
7s_{1/2}^2 7p_{1/2}^1$ excited state of Th$^-$ is $\approx1.5$ and
$\approx1.1$ times bigger than in the $6d_{3/2}^2 7s_{1/2}^2$
ground state of the Th atom. The IC probabilities in the anion
decreases despite the decay via the additional $6d_{3/2}$ or
$7p_{1/2}$ electrons. This counterintuitive result is a
consequence: a) of a decrease in the amplitudes of the $6d_{3/2}$
and $7s_{1/2}$ wave functions near the nucleus due to an increase
in their diffuseness of upon the addition of extra electron, b) of
mutual compensation in the IC probability due to a kinematic
factor, which depends on the energy of the conversion electron in
the continuum as $E_c^{-1/2}$, and the $E_c^{1/4}$ growth of the
amplitudes of the electron wave functions.
\end{abstract}

\pacs{23.20.Nx, 21.10.Tg, 27.90.+b}
\maketitle

\section{Introduction}
\label{sec:Introduction}

The $^{229}$Th nucleus has a unique low-lying isomeric state
$^{229m}$Th($3/2^+,E_{\text{is}}=8.28 \pm 0.17$~eV)
\cite{Seiferle-19}. The dramatic and controversial story of the
experimental studies of this state --- the discovering of the
level
\cite{Kroger-76,Reich-90,Burke-90,Irwin-97,Richardson-98,Utter-99,Tkalya-99-JETPL,Shaw-99,Browne-01,Zhao-12,Peik-13},
the measuring its energy
\cite{Kroger-76,Reich-90,Helmer-94,Beck-07,Wense-16,Seiferle-17,Masuda-19,Seiferle-19},
magnetic and quadrupole moments \cite{Thielking-18}, charge radius
\cite{Safronova-18}, and half-life \cite{Wense-16,Seiferle-17} is
still far from complete. Increasing the accuracy of the
measurements is important for the creation of ultra precise clock
at the nuclear transition of the optical range
\cite{Peik-03,Rellergert-10, Campbell-12, Peik-15}, which in turn
can be used to study the relative effects of the variation of the
fine structure constant and the strong interaction parameter
\cite{Flambaum-06,Litvinova-09,Berengut-09}. Investigations of
this nuclear state are important for the design the laser at the
nuclear transition \cite{Tkalya-11,Tkalya-13}, to control the
isomeric level $\gamma$ decay via boundary conditions
\cite{Tkalya-18-PRL}, to detect the decay of the ground state
sublevel of the nucleus into the isomeric state sublevel in the
muonic atom $^{229}$Th \cite{Tkalya-16-PRA}, to check the
exponentiality of the decay law of at long times \cite{Dykhne-98}
and others.

We know now five possible decay channels of the $^{229m}$Th isomer
--- four processes where the electron shell is involved, and the
alpha decay. It is natural to systematize the first four channels
in the framework of the perturbation theory for the quantum
electrodynamics using the order of relevant Feynman diagrams
\cite{Tkalya-04}.

In the first order of the perturbation theory, this is the process
of the emission of a photon by the nucleus, which for low-energy
nuclear levels is practically unobservable. However, the photon
observation is possible if the $^{229m}$Th isomer is put in the
dielectrics with a large band gap, where the thorium atom becomes
effectively ``ionized'' by the chemical environment. For the first
time, this possibility was indicated in
\cite{Tkalya-00-JETPL,Tkalya-00-PRC}.

Internal conversion (IC) is a second-order process. IC is the main
decay channel of the isomer $^{229m}$Th on the valence shells of
the ground state of the Th atom \cite{Strizhov-91}, on the excited
atomic states of Th \cite{Strizhov-91,Bilous-17}, and on the
Rydberg states \cite{Tkalya-19-PRC-IC_Rydb}. Experimentally, IC
was observed in Refs.~\cite{Wense-16,Seiferle-17,Shigekawa-19}.
Another second-order process, namely, the decay of the $^{229m}$Th
isomer during inelastic scattering by metal conduction electrons,
was considered in \cite{Tkalya-99-JETPL}. The nuclear excitation
by electron transition (NEET) \cite{Morita-73} is also of the
second order process. A detailed theory of NEET is given in
\cite{Tkalya-92,Tkalya-07}. This process can play an important
role as an integral part of the electron bridge.

Electron bridge, a third-order process suggested in
\cite{Krutov-68}, was considered for the decay and excitation of
$^{229m}$Th in Refs.~\cite{Strizhov-91, Tkalya-92-JETPL,
Tkalya-92-SJNP}. Later, this process was thoroughly studied
theoretically (see in
\cite{Kalman-94,Tkalya-96,Porsev-10-PRL,Porsev-10-PRA-1+,Karpeshin-17,Muller-19,Borisyuk-19-PRC}),
and there are high hopes for the effective excitation of the
$^{229}$Th nuclei in ion traps.

The $\alpha$ decay of $^{229m}$Th considered in \cite{Dykhne-96,
Varlamov-97} is an important decay channel. The detection of this
process or accompanying bremsstrahlung \cite{Tkalya-99-PRC} could
be the most reliable proof of both the existence of the
$^{229m}$Th isomer and its excitation by the laser radiation
\cite{Tkalya-96}.

In this paper, we investigate the decay of $^{229m}$Th in the
thorium anion, Th${^-}$. Anions are negative ions created when an
atom gains one or more electrons. Particularly often the anions
are formed from the chemical elements in the groups 17 and 16 of
the Periodic table (F$^-$, O$^{2-}$ and so on). These elements
lack, respectively, one or two electrons with respect to the
complete electronic configuration of a noble gas. However, anions
of the other chemical elements can be formed in various
physicochemical processes, too. Laser plasma is a well-studied
universal source of the negative ions \cite{Sil'nov-07}. The laser
ablation method allows one to get the anions of any chemical
element in the Periodic table. In the laser plasma, anions are
produced at a certain stage of the expansion during cooling of the
plasma bunch after the end of the laser pulse. Typical values of
the negative ion currents are tens to hundreds of microamperes. In
Ref.~\cite{Tang-19}, the Th$^-$ anions were produced via the
pulsed Nd:Y-Al-garnet laser ablation of a thorium metal disk.
Further, the anions were accumulated and cooled via buffer gas
cooling in the ion trap. After that, anions were photodetached by
a tunable dye laser and the outgoing photoelectrons were detected.

It turned out that Th${^-}$ is a stable system with the ionization
potential of about 0.6 eV \cite{Malley-09,Tang-19}. It has at
least a couple of levels connected by a strong electric dipole
transition suitable for laser cooling \cite{Tang-19}. Since the
cooled ions in traps manipulated by laser and the laser ablation
(which produces plasma contained as well as positive and
negatively charged ions), as a method of loading the ion trap, are
currently considered as a promising system for studying of
$^{229m}$Th, the knowledge of the decay channels and the lifetime
of the $^{229m}$Th isomer in anions can be very useful for future
experimental research.

\section{Internal conversion in Th$^-$}
\label{sec:IC}

Until recently, the $6d_
{3/2}^2{}7s_{1/2}^2{}7p_{1/2}^1$~$^4G_{5/2}^{\circ}$ state with
the binding energy of 0.368~eV was considered as the ground state
of the thorium anion \cite{Malley-09}. However, as has been shown
in \cite{Tang-19},
$6d_{3/2}^2{}7s_{1/2}^2{}7p_{1/2}^1$~$^4G_{5/2}^{\circ}$ is an
excited state, and the true ground state of Th${^-}$ is the
configuration $6d_{3/2}^3{}7s_{1/2}^2$~$^4F_{3/2}$ with the
binding energy of 0.6~eV. In addition, there are several strong
electric dipole transitions between the bound levels arising from
configurations $6d_{3/2}^3{}7s_{1/2}^2$ and
$6d_{3/2}^2{}7s_{1/2}^2{}7p_{1/2}^1$. These conclusions have been
reached on the basis of measurements of the ionization potential
for the Thorium anion, and large-scale numerical
multiconfiguration Dirac-Hartree-Fock calculations
\cite{Si-18,Tang-19}.

In the following, we present the analysis of the internal
conversion, which uses the bound state wave functions obtained in
Ref.~\cite{Tang-19}: the $6d_{3/2}^2{}7s_{1/2}^2$ wave function
for the ground state of the Thorium atom (Th), the
$6d_{3/2}^2{}7s_{1/2}^2$ wave function for the ground state of the
anion (Th$^-$), and the $6d_{3/2}^2{}7s_{1/2}^2{}7p_{1/2}^1$ wave
function for the anion excited state (Th$^{-^*}$). Thus, we are
able to compare the internal conversion coefficients (ICC) for the
Th atom and anions, obtained within the same approach. This is
important because different codes can give significantly different
results when one calculates ICC from valence shells at ultra-low
energy nuclear transition (see below Table~\ref{tab:ICC_TBSB}).

The internal conversion coefficients per one electron for the
$E(M)L$ nuclear transition with the energy
$\omega_N=E_{\text{is}}$ were calculated using the formulas
%
%
\begin{eqnarray}
\alpha_{E/ML} = \frac{\omega_N}{m} \frac{E+m}{p} \frac{L}{L+1}
\sum_{f} \left( C^{j_f 1/2}_{j_i 1/2 L 0}\right)^2
|{\cal{M}}^{E/ML}_{if}|^2,
\label{eq:ICC_EML}
\end{eqnarray}
where $m$ is the mass of the electron, $E$ and $p$ are the energy
and momentum of the conversion electron satisfying the condition
$E^2 = m^2 + p^2$ (the system of units is $\hbar=c=1$), $j$ is the
total angular momentum of the electron, $C^{j_f 1/2}_{j_i 1/2 L
0}$ is the Clebsch-Gordan coefficient. The electron matrix
elements in Eq.~(\ref{eq:ICC_EML}) are
%
%
\begin{equation}
\begin{split}
{\cal{M}}^{EL}_{if} &= \int_0^{\infty} h_L^{(1)}(\omega_N a_B x)
[g_i(x)g_f(x)+ f_i(x)f_f(x)]x^2 dx,\\
{\cal{M}}^{ML}_{if} &=\cfrac{\kappa_i+\kappa_f}{L}
\int_0^{\infty} h_L^{(1)}(\omega_N a_B x)[g_i(x)f_f(x)+ \\
&\qquad\qquad {}f_i(x)g_f(x)]x^2 dx.
\end{split}
\label{eq:ME}
\end{equation}
Here  $x=r/a_B$, $a_B$ is the Bohr radius, $h_L^{(1)}(\omega_N a_B
x)$ is the Hankel function of the first kind \cite{Abramowitz-64},
$\kappa = l(l+1)-j(j+1)-1/4$, where $l$ is the orbital angular
momentum of the electron.

One can use the well-known representation for the Clebsch-Gordan
coefficient through the $6j$ symbol \cite{Bohr-98-I}
%
%
\begin{equation}
\left(C^{j_f 1/2}_{j_i 1/2 L 0}\right)^2 =(2l_i+1)(2j_f+1)
\left(C^{l_f 0}_{l_i 0 L 0} \right)^2
\left\{
\begin{array}{ccc}
 l_i & 1/2 & j_i \\
 j_f &  L  & l_f
\end{array}
\right\}^2
\label{eq:CGC}
\end{equation}
in order for the selection rules for the parity and the orbital
angular momentum in Eq.~(\ref{eq:ICC_EML}) to be satisfied
automatically. In this case, the $l_i\rightarrow l_i' = 2j_i-l_i$
substitution should be made in Eq.~(\ref{eq:CGC}) for the magnetic
type nuclear transitions.

In the case we are considering here, the nuclear transition energy
is small and the following conditions are fulfilled: $E_b
<\omega_N \ll 1/a_B$, where $E_b$ is the electron binding energy
in the initial state. Therefore,
Eqs.~(\ref{eq:ICC_EML}--\ref{eq:ME}) take the form
%
%
\begin{eqnarray}
\alpha_{E/ML} &=& e^2 \sqrt{\frac{2m}{\omega_N-E_b}}
\frac{[(2L-1)!!]^2}{(\omega_Na_B)^{2L+1}}\frac{L}{L+1} \nonumber
\\
&&\sum_{f} \left( C^{j_f 1/2}_{j_i 1/2 L 0}\right)^2
|{\text{\textsl{m}}}^{E/ML}_{if}|^2, \label{eq:ICC_EML_2}
\end{eqnarray}
where the new electron matrix elements are
%
%
\begin{equation}
\begin{split}
{\text{\textsl{m}}}^{EL}_{if} &= \int_0^{\infty}
[g_i(x)g_f(x)+ f_i(x)f_f(x)]x^{2-L-1} dx,\\
{\text{\textsl{m}}}^{ML}_{if} &= \cfrac{\kappa_i+\kappa_f}{L}
\int_0^{\infty}[g_i(x)f_f(x)+f_i(x)g_f(x)]x^{2-L-1} dx.
\end{split}
\label{eq:ME_2}
\end{equation}

The energy of the nonrelativistic conversion electron in the
continuum is $E_c=mv^2/2=\omega_N-E_b$, where $v$ is the electron
speed. Thus, the factor $\sqrt{2m/(\omega_N-E_b)}$ in
Eq.~\ref{eq:ICC_EML_2} is equal to $2/v$. We will see below that
this factor significantly ``increases'' the internal conversion
coefficient in the Th atom (whose valence shells have a binding
energy of about 6--7 eV), because it compensates for the small
amplitudes of the electronic wave functions in the continuum.

The  matrix elements (\ref{eq:ME}) and (\ref{eq:ME_2}) were
calculated by numerical integration. We used the wave functions
from the work \cite{Tang-19} for the initial states, and the wave
functions of the continuum for the final state.

\begin{figure}
 \includegraphics[angle=0,width=0.98\hsize,keepaspectratio]{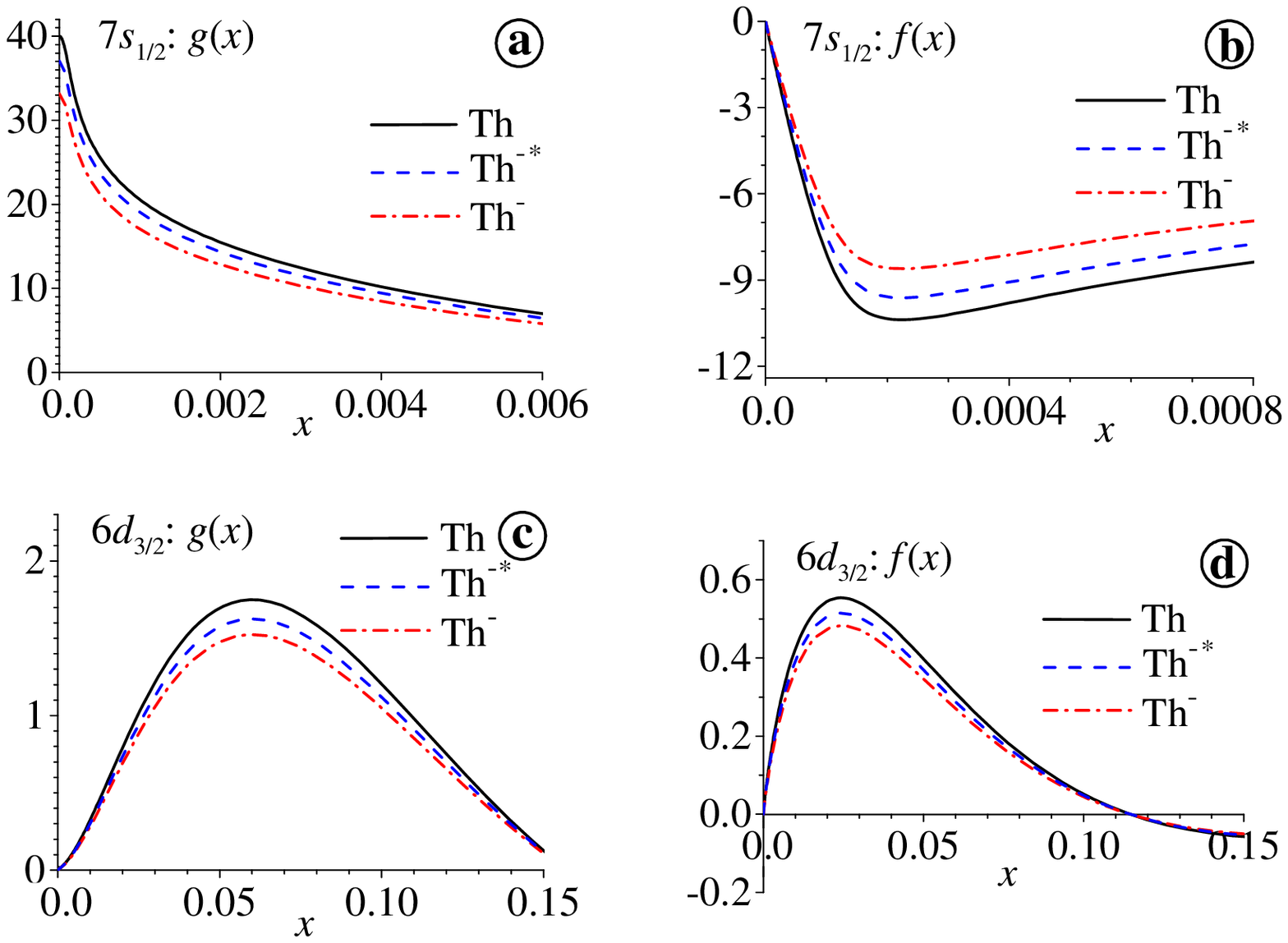}
 \caption{Wave functions of the $7s_{1/2}$ and $6d_{3/2}$ electron states
 in the Th atom and Thorium anion in the ground and excited states: (a) and (c)
 --- the large $g_i(x)$ components, (b) and (d) --- the small $f_i(x)$ components
of the Dirac bispinor.}
 \label{fig:WF}
\end{figure}

\begin{figure}
 \includegraphics[angle=0,width=0.98\hsize,keepaspectratio]{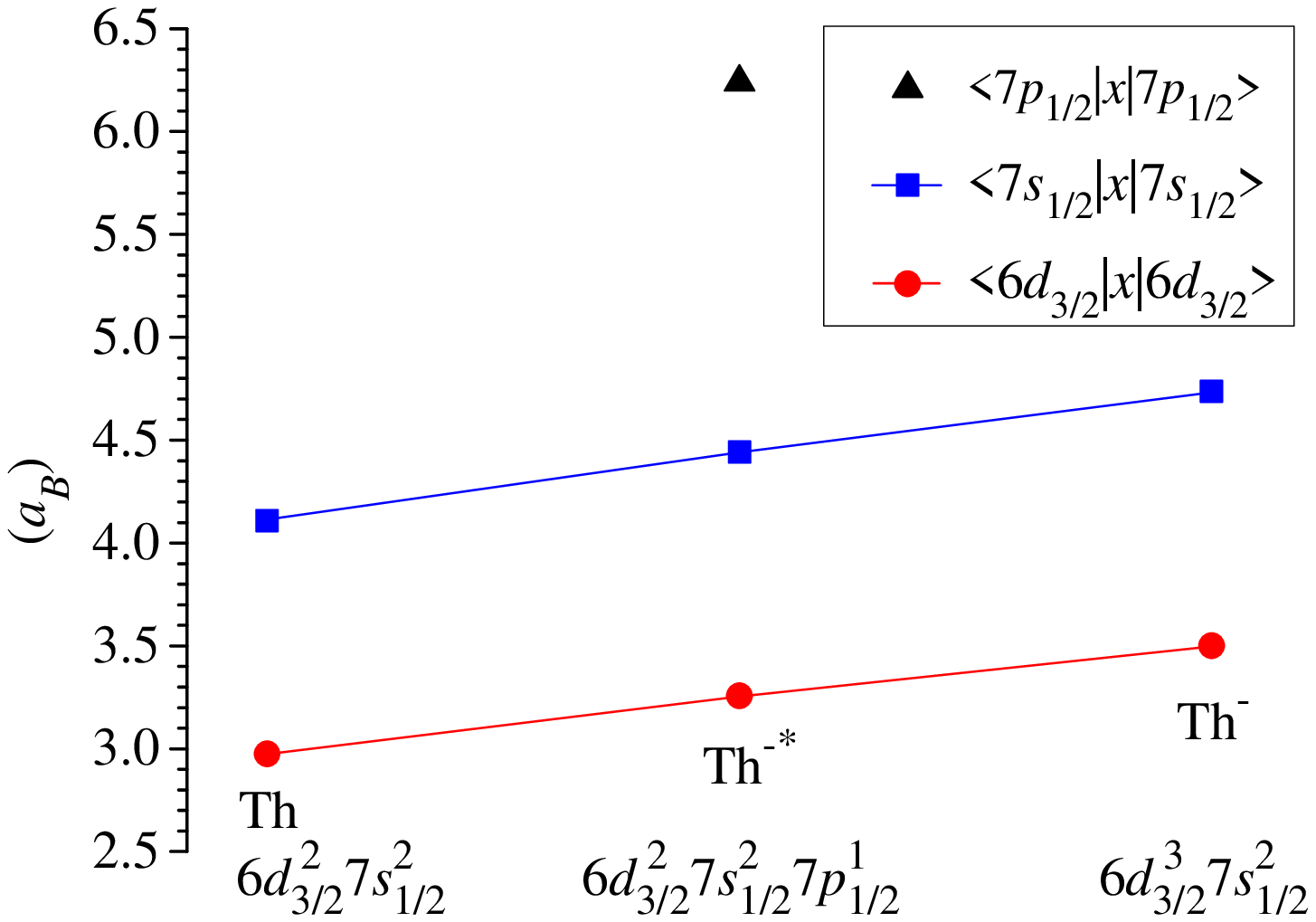}
  \caption{Averaged radii of the $7p_{1/2}$, $7s_{1/2}$ and
$6d_{3/2}$ orbitals in various electron configurations.}
 \label{fig:<x>}
\end{figure}

The wave functions of the initial state are shown partly in
Fig.~\ref{fig:WF} (these regions give main contributions to the
electronic matrix elements).
%
%
Extra electron contributes to an additional nuclear screening for
other valence electrons. As a result, the electron shell becomes
more diffuse. This is clearly seen in Fig.~\ref{fig:<x>} --- the
average orbital radius of the $6d_{3/2}$ and $7s_{1/2}$ states
increases when an electron is added to the $7p_{1/2}$ and
$6d_{3/2}$ states of the Thorium atom. (Note also, in the
$6d_{3/2}^2{}7s_{/2}^2{}7p_{1/2}^1$ anion excited state, the
average radii $\langle{}6d_{3/2}|x|6d_{3/2}\rangle$ and
$\langle{}7s_{1/2}|x|7s_{1/2}\rangle$ are smaller than the
corresponding radii in the $6d_{3/2}^3{}7s_{1/2}^2$ ground state.
It can be easily explained --- the $7p_{1/2}$ shell shields the
nuclear charge less effective than the $6d_{3/2}$ shell because
$\langle{}7p_{1/2}|x|7p_{1/2}\rangle{} >
\langle{}6d_{3/2}|x|6d_{3/2}\rangle$, Fig.~\ref{fig:<x>}.)

%
%

As a result of the indicated ``swelling'', on the one hand, and
the conservation of the normalization volume of the wave
functions, on the other hand, the amplitudes of the $7s_{1/2}$ and
$6d_{3/2}$ wave functions decrease in the region near the nucleus,
Fig.~\ref{fig:WF}.

The wave functions of the continuum spectrum are the numerical
solutions of the Dirac equations with the electron energies $E>m$
($E\approx{}E_c+m$)
%
%
\begin{equation}
\left.
\begin{array}{ll}
g'(x)+\cfrac{1+\kappa}{x}g(x)-
\cfrac{1}{e^2}\left(\cfrac{E}{m}+1-\cfrac{V(x)}{m}\right)f(x) =0,\\
f'(x)+\cfrac{1-\kappa}{x}f(x)+
\cfrac{1}{e^2}\left(\cfrac{E}{m}-1-\cfrac{V(x)}{m}\right)g(x) =0,
\end{array}
\right.
\label{eq:EqDirac}
\end{equation}
normalized at $x\rightarrow\infty$ with the condition
$g_f(x)=\sin(pa_Bx+\varphi_{lj})/x$, where $\varphi_{lj}$ is a
phase. In Eq.~(\ref{eq:EqDirac}), $e$ is the electron charge.

\begin{figure}
 \includegraphics[angle=0,width=0.98\hsize,keepaspectratio]{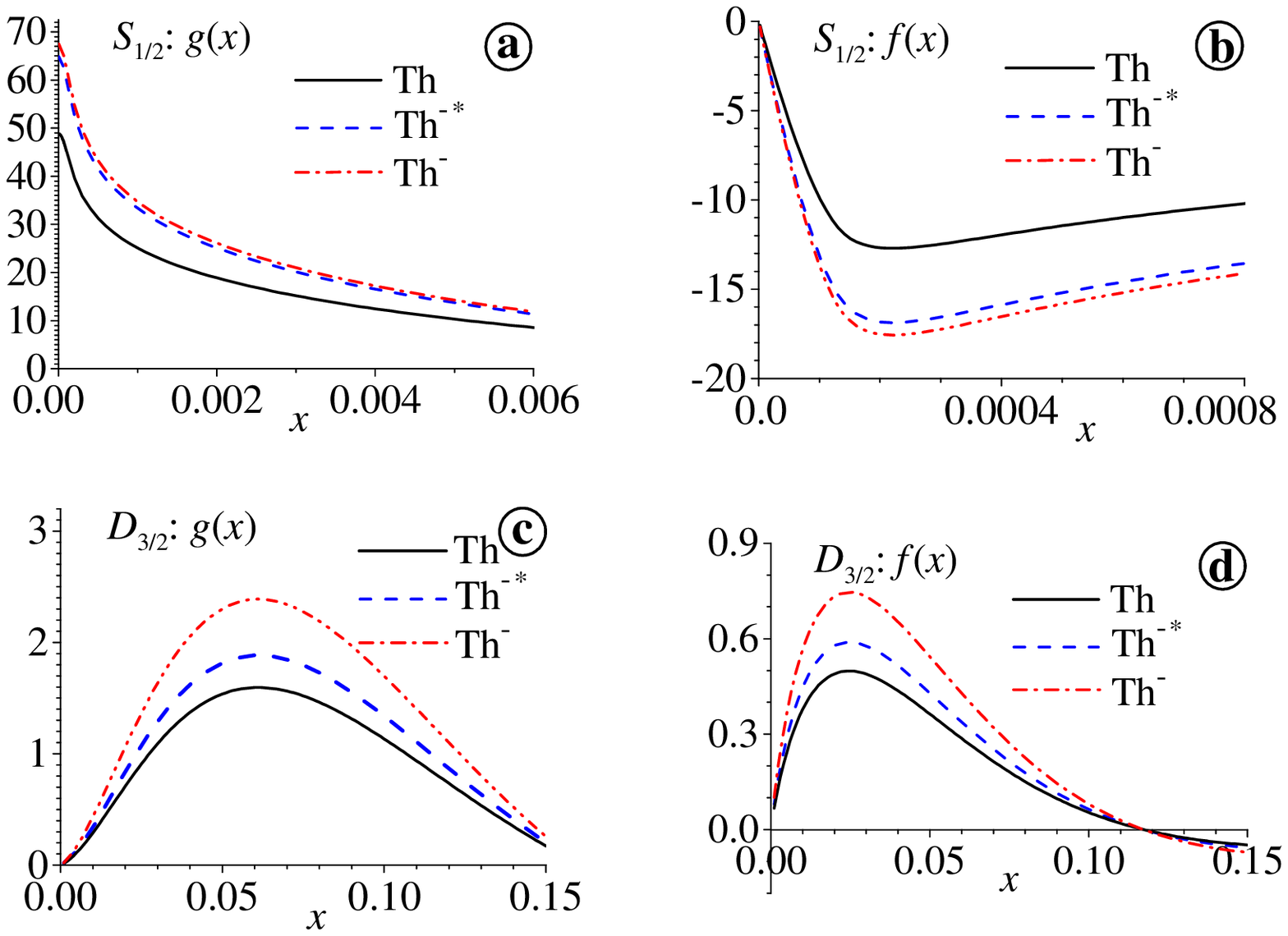}
  \caption{Wave functions of the $S_{1/2}$ and $D_{3/2}$ electron
  states in the continuum after IC on the $7s_{1/2}$ and $6d_{3/2}$ electron states
  in Th, Th$^-$ and Th$^{-^*}$: (a) and (c)  --- the large $g_f(x)$ components,
  (b) and (d) --- the small $f_f(x)$ components of the Dirac bispinor.}
\label{fig:WF-continuum}
\end{figure}

As an example, two wave functions of the final state are shown in
Fig.~\ref{fig:WF-continuum} for the IC transitions
$7s_{1/2}\rightarrow{}S_{1/2}$ and $6d_{3/2}\rightarrow{}D_{1/2}$.
The energies of conversion electrons are: $E_c(S_{1/2})=1.79$~eV
and $E_c(D_{3/2})=1.03$~eV for IC in the Th atom,
$E_c(S_{1/2})=6.68$~eV and $E_c(D_{3/2})=7.65$~eV for IC in the Th
anion in the ground state, and $E_c(S_{1/2})=5.63$~eV and
$E_c(D_{3/2})=5.47$~eV for IC in the Th anion in the excited
state. The solutions $g_f(x)$ (and $f_f(x)$) of
Eq.~(\ref{eq:EqDirac}) reliably reach the asymptotic behavior
$xg_f(x)={\text{Const}}\times{}\sin(pa_Bx)$ at $x\approx300$ in
the Th$^+$ potential (for IC in the Th atom) and at $x\approx30$
in the potential of the Th atom (for IC in Th$^-$ and Th$^{-^*}$).
Further, the obtained wave functions $g_f(x)$ and $f_f(x)$ are
renormalized by dividing by the constant ``Const''.

%
%

The potential energy $V(x)$ of the electron in
Eq.~\ref{eq:EqDirac} is $ V(x)= V_{\text{nucl}}(x) +
V_{\text{shell}}(x)$, where $V_{\text{shell}}(x)$ is the potential
energy of the electron in the shell electron potential, and
$V_{\text{nucl}}(x)$ is the potential energy of electron in
potential of the unscreened nucleus. That is, the positive charge,
$Z$, has been uniformly distributed within a sphere of the radius
$x_{R_0} = R_0/a_B$ ($R_0 = 1.2A^{1/3}$ fm is the radius of the
nucleus with the atomic number $A$): $V_{\text{nucl}}(x) =
-{\cal{E}}_0 (Z/2x_{R_0}) [3-(x/x_{R_0})^2]$ for
$0\leq{}x\leq{}x_{R_0}$, and $V_{\text{nucl}}(x) =-{\cal{E}}_0
Z/x$ for $x\geq{}x_{R_0}$ where ${\cal{E}}_0=me^4$ is the atomic
unit of energy.

The electron shell potential has been found by solving the Poisson
equation with the given electron density. The electron density in
the Th$^+$ ion and in the Th atom for the IC calculations in the
Th atom and in the Th$^-$ anion respectively has been obtained
within the DFT theory \cite{Nikolaev-15,Nikolaev-16} through the
self-consistent procedure taking into account the exchange and
correlation effects. Moreover, for the internal conversion in the
neutral Th atom we consider two various electron densities for
Th$^+$, corresponding to the $6d_{3/2}^2{}7s_{1/2}^1$ and
$6d_{3/2}^1{}7s_{1/2}^2$ configurations.

\section{Results and Discussion}
\label{sec:Discussion}

Calculated internal conversion coefficients are presented in
Table~\ref{tab:ICC}. We estimated the half-life of the $^{229m}$Th
isomer in the anion for the two sets of reduced nuclear
probabilities given in Table~\ref{tab:Bwu}. The first set (see in
Ref.~\cite{Tkalya-15-PRC}) was obtained with Alaga rules from the
available experimental data
\cite{Bemis-88,Gulda-02,Barci-03,Ruchowska-06} for the $M1$ and
$E2$ transitions between the rotation bands $3/2^+[631]$ and
$5/2^+[633]$ in the $^{229}$Th nucleus
\cite{Dykhne-98,Tkalya-15-PRC}). The second set (taken from
Ref.~\cite{Minkov-17}) is based on a detailed computer calculation
using modern concepts of nuclear interactions. The corresponding
probabilities of radiative transitions in the $^{229}$Th nucleus
from the isomeric to the ground state are also given in
Table~\ref{tab:Bwu}.

%
%
\begin{table}
\caption{Internal conversion coefficients per one electron for
nuclear transition with the energy $\omega_N=8.27$~eV for the
Thorium atom in the $6d_{3/2}^2{}7s_{1/2}^2$ ground state (Th),
for the Thorium anion in the $6d_{3/2}^3{}7s_{1/2}^2$ ground state
(Th$^-$) and in the $6d_{3/2}^2{}7s_{1/2}^2{}7p_{1/2}^1$ excited
state (Th$^{-^*}$). (The binding energies on the shells are given
in parentheses).}
 \label{tab:ICC}
 \begin{tabular}{l@{\quad}c@{\quad}c@{\quad}c}
    \hline
    \hline
    Th            & $7s_{1/2}$(-6.49~eV) & $6d_{3/2}$(-7.25~eV) & \\
    \hline
    $\alpha_{M1}$ &$7.93\times10^8$      & $2.31\times10^6$     & \\
    \hline
    $\alpha_{E2}$ &$1.06\times10^{15}$   & $4.80\times10^{15}$  & \\
    \hline
    \hline
    Th$^-$        & $7s_{1/2}$(-1.60~eV) & $6d_{3/2}$(-0.63~eV) & \\
    \hline
    $\alpha_{M1}$ &$5.40\times10^8$      & $1.45\times10^6$     & \\
    \hline
    $\alpha_{E2}$ &$6.59\times10^{14}$   & $3.08\times10^{15}$  & \\
    \hline
    \hline
    Th$^{-^*}$    & $7s_{1/2}$(-2.65~eV) & $6d_{3/2}$(-2.81~eV) & $7p_{1/2}$(-0.61~eV) \\
    \hline
    $\alpha_{M1}$ &$6.76\times10^8$      & $1.21\times10^6$     & $3.37\times10^7$     \\
    \hline
    $\alpha_{E2}$ &$6.85\times10^{14}$   & $2.76\times10^{15}$  & $5.44\times10^{16}$  \\
    \hline
  \end{tabular}
\end{table}
%

%
%
\begin{table}
\caption{Reduced matrix elements ($B_{\text{W.u.}}$) of the
$3/2^+[631](8.27\,\,\text{eV})\rightarrow 5/2^+[633](0.0)$ nuclear
transition in the $^{229}$Th nucleus and corresponding radiation
widths ($\Gamma^{\text{rad}}$).}
 \label{tab:Bwu}
 \begin{tabular}{l@{\quad}c@{\quad}c@{\quad}c}
    \hline
    \hline
    Set & Mult. & $B_{\text{W.u.}}$     & $\Gamma^{\text{rad}}$ (eV) \\
    \hline
    1   & $M1$  &$3.1\times 10^{-2}$    & $3.65\times10^{-19}$       \\
    \cline{2-4}
        & $E2$  &$11.7$                 & $3.06\times10^{-29}$       \\
    \hline
    2   & $M1$  &$0.76\times10^{-2}$    & $8.94\times10^{-20}$       \\
    \cline{2-4}
        & $E2$  &$27$                   & $7.05\times10^{-29}$       \\
    \hline
\end{tabular}
\end{table}

With data presented in Tables~\ref{tab:ICC} and \ref{tab:Bwu} we
calculate the half-life $T_{1/2}$ of the isomer $^{229m}$Th in the
atom and anion. The results are summarized in
Table~\ref{tab:T_1/2}. They must be treated with some caution. The
accuracy of calculating the IIC is relatively small for the
nuclear transitions with ultralow energies (see below in
Table~\ref{tab:ICC_TBSB}). This is mainly due to the accuracy of
the calculation of the wave functions of valence states and their
binding energies. That is why we have calculated the internal
conversion probabilities not only for the thorium anion, but also
for the thorium atom. Since the calculations have been performed
in a unified approach, we consider these results as reliable for
the description of relative changes in the conversion
probabilities and half-lives of $^{229m}$Th in going from atom to
anion.

As can be seen from Table~\ref{tab:T_1/2}, the lifetime of the
isomer in the $6d_{3/2}^3{}7s_{1/2}^2$ ground state of the anion
is approximately 1.4--1.5 times longer than the isomer lifetime in
the atom. For the Thorium anion in the
$6d_{3/2}^2{}7s_{1/2}^2{}7p_{1/2}^1$ excited state this excess is
insignificant, only $\approx$10\%.

%
%
\begin{table}
\caption{$^{229m}$Th isomer half life (in s) in the Th atom and in
the Thorium anion in the ground (Th$^-$) and in the excited
(Th$^{-^*}$) states.}
 \label{tab:T_1/2}
 \begin{tabular}{l@{\quad}c@{\quad}c@{\quad}c@{\quad}c}
    \hline
    \hline
    Set &                               & Th                    & Th$^-$                & Th$^{-^*}$            \\
    \hline
    1   & $T_{1/2}$                     & $7.86\times 10^{-7}$  &$1.15\times 10^{-6}$   & $8.98\times10^{-7}$   \\
    \cline{2-5}
        &$T_{1/2}/T_{1/2}^{\text{Th}}$  & 1                     &1.47                   & 1.14                  \\
    \hline
    2   & $T_{1/2}$                     & $3.19\times 10^{-6}$  &$4.67\times10^{-6}$    & $3.55\times10^{-6}$   \\
    \cline{2-5}
        &$T_{1/2}/T_{1/2}^{\text{Th}}$  & 1                     &1.47                   & 1.11                  \\
    \hline
\end{tabular}
\end{table}

\begin{figure}
 \includegraphics[angle=0,width=0.98\hsize,keepaspectratio]{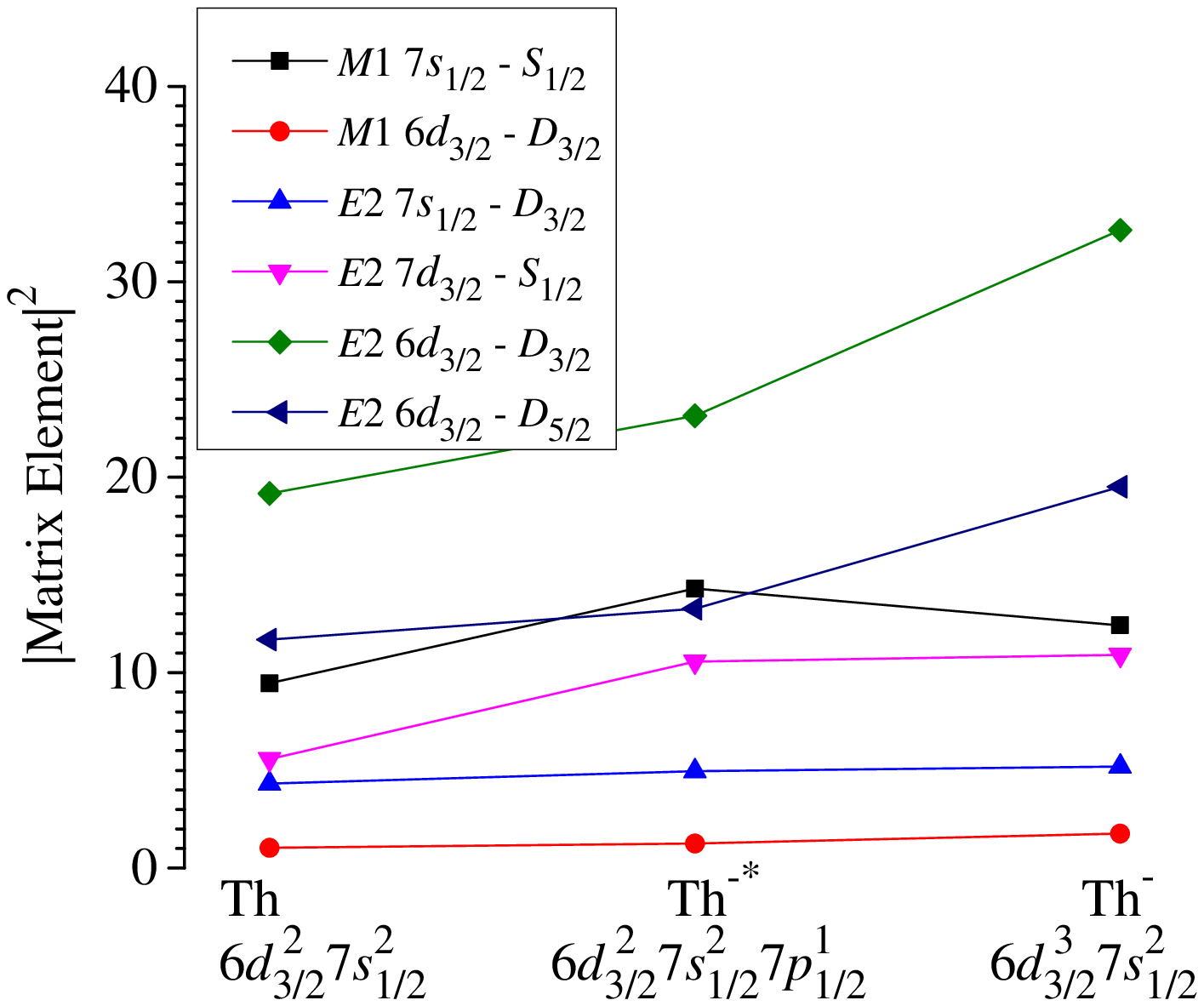}
  \caption{Matrix Elements Eq.~(\ref{eq:ME_2}) for the main IC transitions.
  (Matrix elements for the $M1$ $6d_{3/2}$\,-$D_{3/2}$ electronic
  transitions are given in the units of $10^{-1}$.)}
  \label{fig:ME2}
\end{figure}

This result at first glance looks counterintuitive. First, the
internal conversion proceeds on four valence electrons in the
thorium atom, and on five valence electrons in the anion. Second
-- less obvious -- reason is that the electron matrix elements in
the anion are larger than in the Th atom (see in
Fig.~\ref{fig:ME2}). (Note that the latter effect is nontrivial,
since, as we have seen, the WF amplitudes of the bound $6d_{3/2}$
and $7s_{1/2}$ states decrease upon transition from the Th atom to
the thorium anion. The matrix elements growth in
Fig.~\ref{fig:ME2} is explained by a faster increase in the
amplitudes of the electron wave functions in the continuous
spectrum with an increase of its energy (see in
Fig.~\ref{fig:WF-continuum}). Due to diffusion of the electron
shell, the binding energies of the electrons in the Thorium anion
are smaller than in the atom, and the kinetic energy of the
conversion electrons in the continuum is greater. In the
nonrelativistic case, the amplitudes of the Coulomb wave functions
(see in \cite{Abramowitz-64}) in the continuum increase with the
energy of the conversion electron as $E_c^{1/4}$, i.e.
significantly faster than the decrease of the amplitudes of the
wave functions in the discrete spectrum. This explains somewhat
unexpected form of the plots in Fig.~\ref{fig:ME2}.)

%
%

There is also a third factor that places an important role in the
process. This is the kinetic energy of the conversion electron in
the denominator in Eq.~(\ref{eq:ICC_EML_2}). It gives the factor
$1/v$ in the expression for the IC probability and increases it
near the threshold of the Th atom. In the thorium anion, this
factor is 2-3 times smaller. And such a decrease turns out to be
the most significant reason, which compensates for the increase in
amplitudes of the electron wave functions in the continuum, and
leads to a decrease in the IC probability of the anion in both the
ground and excited states.

It is necessary to emphasize one more feature of the internal
conversion in the anion. According to Fig.~\ref{fig:<x>},
$\langle{}6d_{3/2}|x|6d_{3/2}\rangle$ is the largest in the ground
state of the anion. Nevertheless, the partial internal conversion
coefficients on the $6d_{3/2}$ shell in Th$^-$ exceed ICC in
Th$^{-^*}$ (see Table~\ref{tab:ICC}). This is caused by the lack
of the cancellation of two effects: the $E_c^{1/4}$ increase in
the amplitudes of WF of the conversion electron in the continuum
and the decrease of the factor $1/v$ for the electron promoted
during IC from the $6d_{3/2}$ shell of Th$^-$. We recall that we
use the energies for the bound states of the electron orbitals
from Ref.~\cite{Tang-19}, where the multi-configuration electron
terms were taken into account. In this case, the binding energies
of the terms shift up or down, while the radial wave function
remains the same. As a result, one gets the indicated
inconsistency between the amplitude of the radial wave function
and the binding energy of the orbital. Note that this effect
practically does not affect the main results of the work.

In conclusion, it will be useful to compare the total IC
coefficients for the $7s_{1/2}$ and $6d_{3/2}$ shells of the Th
atom obtained in different works and using different codes. The
relevant data are given in Table IV. We see that on average all
the data correspond to each other with the accuracy of the factor
of two. This is sufficient for preliminary estimates of the isomer
lifetime and planning of experiments. For more delicate effects,
it is necessary to perform calculations within the same code.

%
%
\begin{table}
\caption{Total IC coefficients for the $7s_{1/2}$ and $6d_{3/2}$
shells of the Th atom obtained by different codes for the two
values of the isomeric level energy: $E_{\text{is}} = 8.28$~ eV,
and 7.8~eV. (In the parentheses, there are the binding energies on
the shells.)}
 \label{tab:ICC_TBSB}
\begin{tabular}{|l|c|c|c|c|}
    \hline
    \hline
    This & \multicolumn{2}{|c|}{$7s_{1/2}$(-6.49~eV)} & \multicolumn{2}{|c|}{$6d_{3/2}$(-7.25~eV)}\\
    \cline{2-5}
    work & \multicolumn{1}{|c|}{$E_{\text{is}}=8.28$~eV}&
                    \multicolumn{1}{|c|}{$E_{\text{is}}=7.8$~eV}&
                                    \multicolumn{1}{|c|}{8.28~eV}&
                                            \multicolumn{1}{|c|}{7.8~eV}\\
    \hline
    $\alpha_{M1}$ &$1.6\times10^9$      &  & $4.6\times10^6$ &    \\
    \hline
    $\alpha_{E2}$ &$2.1\times10^{15}$   & &  $9.6\times10^{15}$ &  \\
    \hline
    \hline
    Ref.~\cite{Bilous-18} & \multicolumn{2}{|c|}{$7s_{1/2}$(no data)} & \multicolumn{2}{|c|}{$6d_{3/2}$(no data)}\\
    \cline{2-5}
     & \multicolumn{1}{|c|}{$E_{\text{is}}=8.28$~eV}&
                    \multicolumn{1}{|c|}{$E_{\text{is}}=7.8$~eV}&
                                    \multicolumn{1}{|c|}{$8.28$~eV}&
                                            \multicolumn{1}{|c|}{$7.8$~eV}\\
    \hline
    $\alpha_{M1}$ &   & $1.1\times10^9$    &  & $2.0\times10^6$    \\
    \hline
    $\alpha_{E2}$ &   & $4.8\times10^{15}$&  & $4.3\times10^{15}$ \\
    \hline
    \hline
    Ref.~\cite{Tkalya-15-PRC} & \multicolumn{2}{|c|}{$7s_{1/2}$(-5.20~eV)} & \multicolumn{2}{|c|}{$6d_{3/2}$(-4.20~eV)}\\
    \cline{2-5}
    code\cite{Soldatov-79} & \multicolumn{1}{|c|}{$E_{\text{is}}=8.28$~eV}&
                    \multicolumn{1}{|c|}{$E_{\text{is}}=7.8$~eV}&
                                    \multicolumn{1}{|c|}{$8.28$~eV}&
                                            \multicolumn{1}{|c|}{$7.8$~eV}\\
    \hline
    $\alpha_{M1}$ & $1.6\times10^9$  & $1.9\times10^9$  & $4.1\times10^6$ &  $4.9\times10^6$  \\
    \hline
    $\alpha_{E2}$ & $2.1\times10^{15}$  & $2.8\times10^{15}$ & $9.4\times10^{15}$  & $1.3\times10^{16}$  \\
    \hline
    \hline
    Ref.~\cite{Borisyuk-18-QE} & \multicolumn{2}{|c|}{$7s_{1/2}$(-5.62~eV)} & \multicolumn{2}{|c|}{$6d_{3/2}$(-6.10~eV)}\\
    \cline{2-5}
    code\cite{Band-79} & \multicolumn{1}{|c|}{$E_{\text{is}}=8.28$~eV}&
                    \multicolumn{1}{|c|}{$E_{\text{is}}=7.8$~eV}&
                                    \multicolumn{1}{|c|}{$8.28$~eV}&
                                            \multicolumn{1}{|c|}{$7.8$~eV}\\
    \hline
    $\alpha_{M1}$ & $0.96\times10^9$  & $1.1\times10^9$  & $2.8\times10^6$ &  $3.3\times10^6$  \\
    \hline
    $\alpha_{E2}$ & $1.2\times10^{15}$ & $1.6\times10^{15}$ & $6.0\times10^{15}$  & $8.0\times10^{15}$  \\
    \hline
  \end{tabular}
\end{table}

\section{Conclusion}
\label{sec:Conclusion}

In the paper, for the first time, we have studied the decay of the
low lying isomer $3/2^+(8.28\pm0.17$ eV) of the $^{229}$Th nucleus
in the Thorium anion. It has been found that the half life of the
isomer in the $6d_{3/2}^3{}7s_{1/2}^2$ ground state of the anion
is approximately 40-50{\%} above the value in the
$6d_{3/2}^2{}7s_{1/2}^2$ ground state of the Th atom and
$\approx$10{\%} larger in comparison with the
$6d_{3/2}^2{}7s_{1/2}^2{}7p_{1/2}^1$ excited state of the anion.
The IC decay probability is correspondingly reduced, despite
``extra'' fifth electron involved in the internal conversion
process. The reason is as follows. Extra electron contributes to
an additional nuclear screening for other valence electrons. As a
result, the valence electron shells become more diffuse and
amplitudes of the $6d_{3/2}$ and $7s_{1/2}$ wave functions near
the nucleus decrease. Following the amplitudes, the probability of
the internal conversion decreases too.

This research was supported by a grant of the Russian Science
Foundation (Project No 19-72-30014).

\end{document}